\documentclass[twocolumn]{aa}
\usepackage{graphicx}
\begin{document}

\title{Evidence for a current sheet forming in the wake of a Coronal Mass Ejection from multi-viewpoint
coronagraph observations}

 \author{S. Patsourakos,
          \inst{1} and
         A. Vourlidas\inst{2} }
          
 \institute{University of Ioannina, Department of Physics, 
Section of Astrogeophysics, Ioannina, GR-45110, Greece\\
              \email{spatsour@cc.uoi.gr}
         \and
             Space Science Division, Naval Research Laboratory, Washington, DC 20375, USA \\
             \email{vourlidas@nrl.navy.mil} 
}
  \date{}

  \abstract
  {Ray-like features observed by coronagraphs in the wake of Coronal
    Mass Ejections (CMEs) are sometimes interpreted as the white light
    counterparts of current sheets (CSs) produced by the eruption.
    The 3D geometry of these ray-like features is largely unknown and its
    knowledge should clarify their association to the CS and place
    constraints on CME physics and coronal conditions.}
{If these rays are related to field relaxation behind CMEs, therefore
  representing current sheets, then they should be aligned to the CME
  axis. With this
  study we test these important implications for the first time.  }
{An example of such a post-CME ray was observed by various coronagraphs,
  including these of the  Sun Earth Connection Coronal and Heliospheric
investigation (\textsl{SECCHI}) onboard the
Solar Terrestrial Relations Observatory
(\textsl{STEREO}) twin spacecraft
and  the Large Angle Spectrometric Coronagraph (\textsl{LASCO}) onboard
the Solar and Heliospheric Observatory (\textsl{SOHO}). The ray
was observed 
 in the aftermath of a CME which occurred
  on 9 April 2008.  The twin STEREO spacecraft were separated by about
  $48^\circ$ on that day.  This significant separation combined with
  a third "eye" view supplied by LASCO allow for a truly multi-viewpoint
  observation of the ray and of the CME.  We applied 3D forward
  geometrical modeling to the CME and to the ray as simultaneously
  viewed by SECCHI-A and B and by SECCHI-A and LASCO, respectively. }
{We found that the ray can be approximated by a rectangular slab,
  nearly aligned with the CME axis, and much smaller than the CME in
  both terms of thickness and depth ($\approx$ 0.05 and 0.15
  $R_{\odot}$ respectively). The ray electron density and temperature were substantially
  higher than their values in the ambient corona. We found that the 
ray and CME are significantly displaced from the associated post-CME flaring loops.}
{The properties and location of the ray are fully consistent with the
  expectations of the standard CME theories for post-CME current
  sheets. Therefore, our multi-viewpoint observations supply strong
  evidence that the observed post-CME ray is indeed related to a
  post-CME current sheet.}

   \keywords{Sun: coronal mass ejections (CMEs) -- Sun: flares}
\authorrunning{Patsourakos and Vourlidas}
\titlerunning{STEREO SOHO observations of a post-CME current sheet proxy}

\maketitle

\section{Introduction}\label{intro}

CMEs result from a major restructuring of the corona.  During such
events, magnetic fields and plasma are expelled from the Sun causing
dramatic changes in the coronal medium both within and around the
erupting volume.

Essentially all CME models predict
the formation of a post-CME  current sheet
(CS)
underneath the erupting magnetic flux rope (e.g.
see also the review of  Forbes and Lin 2000). 
As
the magnetic flux rope \footnote{Flux ropes are currently
widely recognized as the major constituent
of CMEs.} ascends, 
it pushes aside
overlying  magnetic fields.
These fields will pinch off
underneath the erupting rope forming
a  neutral CS, i.e.
a surface separating fields
of opposite magnetic polarity.
The faster the eruption  proceeds, the higher the
rate at which new field lines accumulate
at the CS. Such post-eruption 
CSs should be distinguished from
the CSs which could exist {\it before} the eruption
in some CME models.

There exist several pieces of observational evidence supporting the
existence of post-CME CSs both in the inner and in the outer corona. A
major "smoking gun" for their existence is in the form of White Light
(WL) rays seen by coronagraphs behind CMEs (e.g., 
Ko et al. 2003;
Webb et
al. 2003; Vr{\v s}nak et al. 2009).  The rays are formed
in the wake of CMEs, they extend to several solar radii and persist
from several hours up to few days.  The detection of very hot ($>3$
MK) plasmas in several of these WL rays by
the UltraViolet Coronagraph Spectrometer
(UVCS; Kohl et al. 1995)  further supported the
possibility that such structures represent CSs (e.g., Ko et al. 2003;
Raymond et al. 2003; Bemporad et al. 2006; Lee et al. 2006; Lin et
al. 2007).  Coronal CSs are places where one could expect higher
heating rates, and therefore higher temperatures. Moreover, such rays are not quiescent but they often
exhibit dynamics such as propagating blobs through the ray (e.g., Sheeley and Wang
2007).  Such dynamic phenomena could result from reconnections taking
place in a CS.  A recent review of CS observations made by LASCO and
UVCS is given in Vr{\v s}nak et al. (2009).  An analytical
Petschek-like current sheet model was found in a good agreement with
the observations.  Closer to the solar surface, observations of
cusp-like hot structures and descending voids above eruption sites and
arcades provided further support for the existence of post-CME CSs
(e.g., McKenzie and Hudson 1999; Innes et al. 2003; Asai et al. 2004;
Sheeley et al. 2004).

However there are a couple of reasons which argue against the
possibility the observed rays could represent CSs. First, standard
plasma theory predicts that  CSs should
be very narrow (a few meters or less) for typical coronal conditions
 (e.g., Litvinenko 1996; Wood 
\& Neukirch 2005). However, the coronagraph
observations suggest CS widths at a fraction of the solar radius 
$0.2-0.8 R_{\odot}$: Vr{\v s}nak et al. 2009).
One way to overcome this apparent contradiction is to invoke anomalous
processes (e.g., turbulence, hyperesistivity) which could
substantially broaden CSs (e.g., Lin et al. 2007; Bemporad 2008).
Another possibility is that the macroscopic CS could be in reality the
superposition of several small hot-spots, patches where small-scale
reconnections take place (e.g., Klimchuk 1996; Cargill et al. 2006; Linton and
Longcope 2006; Onofri, Isliker, 
\& Vlahos, L.\ 2006, Vr{\v s}nak et al. 2009).

A second difficulty comes from
the very nature of all past ray observations
which employed a single view
point. This does not allow 
to determine the full 3D geometry (i.e., width
depth and height) of the
rays and their exact relationship  
with the associated CMEs.
All we could measure are 
quantities projected in the plane of sky (POS) of the
corresponding instrument.
Even more importantly one
has to worry that projection effects
in the optically thin corona
could cause one to falsely believe that the observed
rays represent  post-eruption
CSs. This is particularly true 
 during CMEs
which cause reconfigurations  of significant  coronal volumes
in terms of rotations, deflections, displacements etc
of structures both within and around the CME. 
For instance the outer CME legs or structures around
the erupting volume, could
be falsely identified as candidate CSs when observed
at a favorable angle;  they may appear
to trail the CME in projection as viewed from a single viewpoint.
However, such structures
may be far from the eruption core where the CS should
lie.

Therefore we do not know the "true"
width and depth of these rays, not to mention
their shapes.
Moreover, we currently
ignore how they are
arranged in 3D space with respect to
the associated CME and whether they represent
a projection effect. From the standard
theory for the formation of post-CME
CSs one would expect them to 
be quasi-aligned with the  CME and to be bounded within it
(i.e., to occupy a smaller volume than the CME itself).
Clearly observations of post-CME rays from multiple
and distinct viewpoints
are required in order to address the above
important questions.

The launch of the STEREO mission in late 2006 initiated an avenue for
the study of the solar corona in 3D.  The mission consists of two
almost identical spacecraft, one orbiting ahead (STA) and the other
orbiting behind (STB) Earth's orbit. The two spacecraft drift apart at
a rate of $\approx 45^\circ$ per year. With this paper we supply the
first detailed 3D analysis of a post-CME ray using STEREO and LASCO data. Our analysis supplies some new and
important geometrical and physical evidence that the observed ray
could represent a CS.

Our paper is structured as follows.  An overview of the CME-ray
observations is given in Section \ref{over}. Section \ref{3d}
describes  the 3D geometrical modeling applied to the CME and to
the ray whereas in Section \ref{den} we determine various physical
parameters (density and temperature) of the ray. Finally, Section
\ref{disc} summarizes our findings and discusses their implications.

\section{Overview of the Observations}\label{over}

We analyze here a CME event which took place on 9 April 2008.  The
event was initiated in the active region (AR) 1098 of the NOAA
classification.  There was no GOES flare associated with this CME:
this could be due to the fact that the source active region was
partially occulted behind the West limb as seen from the Earth.  The
event was observed by the Solar and Heliospheric Observatory (SOHO), Hinode, 
the Transition Region and Corona Explorer
(TRACE) and STEREO. A synopsis of
these observations was presented by Reeves at al. (2008),  Landi et
al. (2010) and Savage et al. (2010).  

We have posted  a movie (movie1.mpg) showing
the low coronal development of the event.
This movie consists of composite  EUVI-195 \AA \, channel and COR1 Extreme Ultra-violet and WL images respectively
taken by STA; a couple of representative snapshots
from this movie is given in Figure \ref{fest2}; the FESTIVAL
software of Auch{\`e}re et al. (2008) was used
to generate the composite images.

The event
started around 09:10 UT with the activation and the lift off of a
prominence. The eruption moved initially parallel to the surface towards the southeast and was diverted outwards along a southwestern path. Such deflection may have been caused by the
large equatorial coronal hole south of the host active region. 
The strongly non-radial propagation of the early CME can also be evidenced in the upper panel of Figure \ref{fest2}
where we note a significant displacement between the axis of the   emerging CME into the COR1 field of view (FOV)
and the position angle of the host AR.
The
erupting prominence eventually gave rise to a typical 3-part CME (e.g., upper panel
of Figure \ref{fest2}).  A  rather faint linear
feature emitting in soft Xrays in the trail of the erupting CME was
observed by the  X-ray Telecope (XRT) on Hinode 
(Reeves et al. 2008; Savage et al. 2010). Combined analysis of the XRT and 
the EUV Imaging Telescope (EIS) Hinode
observations showed that  this feature had temperatures in the range $\approx$
6--10 MK (Landi et al. 2010).  
This ultra-hot linear
structure in the wake of a CME may  have been associated with a 
post-CME CS.  Indeed, 
XRT count-rates as predicted from a Petchek-like current sheet
model were found in a good agreement with the
weak signal of the mentioned above current-sheet like
feature (Ko et al. 2010).
Once the CME exited the COR1 FOV a WL ray was formed
at its wake (e.g., lower panel of Figure \ref{fest2}).
Also Savage et al. (2010) observed with XRT mass flows
(e.g., outflows, supra-arcade downflows) around the mentioned above
faint linear feature which supplied evidence that this structure possibly
represented a post-CME CS.

We concentrate here on WL observations of the corona taken by the COR2
coronagraphs (FOV from 2.5-15 $R_{\odot}$) of the
Sun-Earth Connection Coronal and Heliospheric Investigation (SECCHI;
Howard et al. 2008) instrument suite on-board the twin STEREO
spacecraft and by the C2 coronagraph (FOV from 2.5-6 $R_{\odot}$) of
the LASCO experiment on SOHO (Brueckner et al. 1995).  On the day of our
observations the twin STEREO spacecraft were separated by $\approx$ 48$^\circ$.  Our combined multi-viewpoint observations allowed to
monitor and characterize in 3D the evolution and the structure of the
CME and of the post-CME ray.  Raw data were first treated with the
standard data reduction routines.  COR2 obtains both total WL
brightness (TB) and polarized WL brightness (PB) images, whereas C2
normally obtains only TB images.  Typical image cadence is of the
order of 10-20 minutes.  There were no UVCS observations of the ray as
the instrument was observing lower in the corona (Landi et al. 2010).

\begin{figure*}[p]
\includegraphics[scale=.5,angle=90]{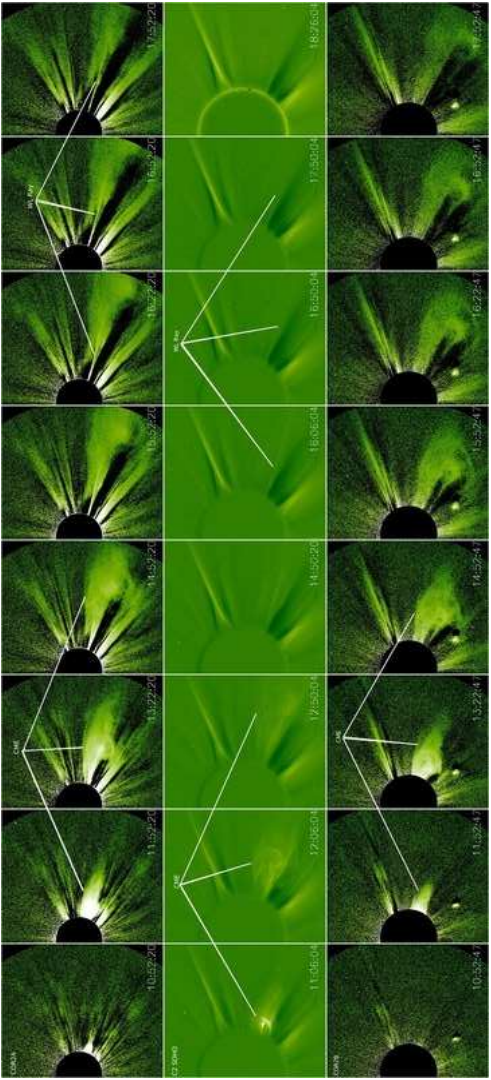}
\caption{Representative total brightness (TB) snapshots of the CME and of the WL ray as observed by COR2A, LASCO C2 and COR2B (upper, middle
and lower panel respectively). 
Pre-event TB images taken at 01:27:13 UT (C2) and 09:22:00 UT (COR2A-B) were subtracted from each C2, COR2A, and COR2B frame. 
The C2 field of view spans 2.2-6 $R_{\odot}$ and the COR2A-B view spans 2.5-15 $R_{\odot}$. 
The temporal evolution of the top, middle and bottom panels is shown in movie2.mpg, movie3.mpg and movie4.mpg, respectively.}
\label{contall}
\end{figure*}

Figure \ref{contall} contains
representative TB images throughout  the event for COR2 on STA (COR2A),
C2 and COR2 on STB (COR2B) respectively; the full
sequences can be found in the corresponding
online movies (movie2.mpg, movie3.mpg and movie4.mpg).
From each image in these figures and the associated movies
a pre-event TB image was first subtracted creating
a base-difference image sequence. Therefore,  bright (dark) areas correspond
to mass  increase (decrease) respectively.

Several remarks can now be made. The CME has already emerged
into the COR2A and C2 FOV around 11:00; this happens later for COR2B
and hints to a CME being closer to the STA and SOHO plane-of-sky (POS).
Once it appears in the coronagraph FOVs the CME has the  canonical
3-part CME structure with a bright front, dark cavity and  bright core, i.e. it is
a typical flux-rope CME. The CME core exhibits significant fine structure and consists
of many threads as can be seen better in the C2 images. The CME  seems
to undergo a rotation as it expands. 
Indeed, Thompson, T{\"o}r{\"o}k $\&$ Kliem (2010)
studied the rotation of the erupting prominence that was associated with 
our CME. Using triangulations of several threads of the erupting prominence observed
by STA and STB they determined its rotation profile for heliocentric distances $\approx$ 1.5-3.3 $R_{\odot}$. 
They found that the prominence rotated by $\approx\!120^\circ$ relative to its preeruption position.
The CME starts to exit the  SOHO FOV
around 13:00 UT and the STA and STB FOVs around  15:52 UT and 
16:52 UT, respectively. While this happens we see a WL ray
forming in the wake of the CME in C2 
(e.g., from 14:50 UT onwards in Figure \ref{contall})
and in
COR2A (e.g., from 15:52 UT onwards in Figure \ref{contall});
there is no evidence for such a ray in STB though. The  WL ray seems
to undergo a rotation as the CME itself is rotating during its expansion.
The WL ray connects to the bottom part of the CME cavity
and seems to be formed when linear features
joining the CME core start to pinch off as the
CME propagates outwards.
The  
ray was visible for at least one day after its formation.

From the above we conclude
that the observed WL ray 
could be a candidate CS. 
One needs to always keep in mind that strictly speaking
a CS is a surface separating magnetic fields
of opposite direction: this is something
which can not be determined by our observations
or by any other similar observation given
the lack of direct magnetic field
observations in the corona. The
3D analysis of this paper will supply rather
strong  indirect evidence that the
observed WL ray could have been a post-CME CS.

\begin{figure*}[p]
\includegraphics[scale=.45]{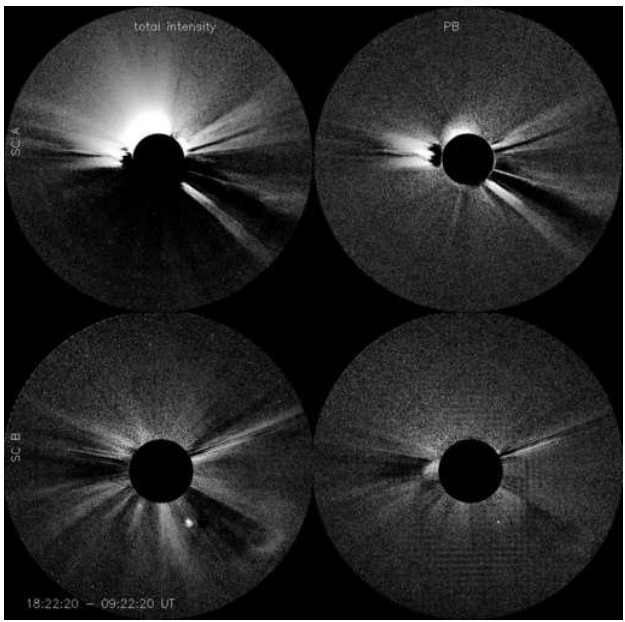}
\caption{TB and PB images for STA (first row) and
for STB (second row).  They were taken at 18:22:20 UT.
Pre-event TB and PB images were subtracted. Field of view spans
from 2.5-15 $R_{\odot}$.}
\label{pb}
\end{figure*}

A first set of clues about the line of sight (LOS) extent of the 
post-CME ray and of the  CME
could be drawn from the analysis
of the  COR2 TB and PB images shown in Figure \ref{pb}.
The theory of Thompson scattering 
(Billings 1966) which describes
the formation of the WL emissions
observed by coronagraphs
predicts that PB decreases
very fast when the emitting electrons lie away 
(i.e., $\approx \pm 20^\circ$)
from the POS of a given instrument (Vourlidas and Howard 2006).
Therefore PB tends to emphasize structures close
to a given POS; this is not the case for TB which
has a more shallow fall-off with distance from the POS.

What we note from Figure \ref{pb} is that the ray
is absent in the STB PB and TB images whereas
it can be seen in STA PB and TB. This means
that the ray should lie close to the  POS
of STA.

Moreover, the LOS extent (depth) of the ray cannot be too large (i.e.,
comparable to the CME width). Otherwise, the ray would be visible in
STB in either TB or PB since the CME itself was seen by both spacecraft in TB
(Figure \ref{contall}) and in PB (Figure \ref{pb}; although less
pronounced in STB).  The above observations supply a rather strong
indication that the ray could not have been due to a projection effect
(e.g, widely separated CME legs lining up along the LOS) but it is
rather a structure with limited extent along the LOS. This is further
demonstrated and quantified in the next Section.

\section{3D Geometrical Modeling of the CME and of the post-CME ray}\label{3d}

In order to determine the 3D structure of the post-CME ray
and to infer its 3D position with respect
to the CME we performed 3D geometrical modeling
of both the CME and the ray. 

We first start with the CME modeling. For this task we used
the 3D forward model of Thernisien et al. (2006, 2009)
(\verb rtsccguicloud.pro  \,  and \verb rtraytracewcs.pro \,
$ssw$ routines). The model has been
successful in fitting the 3D morphology of CMEs (Thernisien et
al. 2006, 2009) observed either from one viewpoint (LASCO) or
from two viewpoints (STA and STB). The latter approach deemed particularly
powerful in strongly constraining the CME parameters.
We used TB images from COR2 on STA and STB.

\begin{figure*}[p]
\includegraphics[scale=0.8]{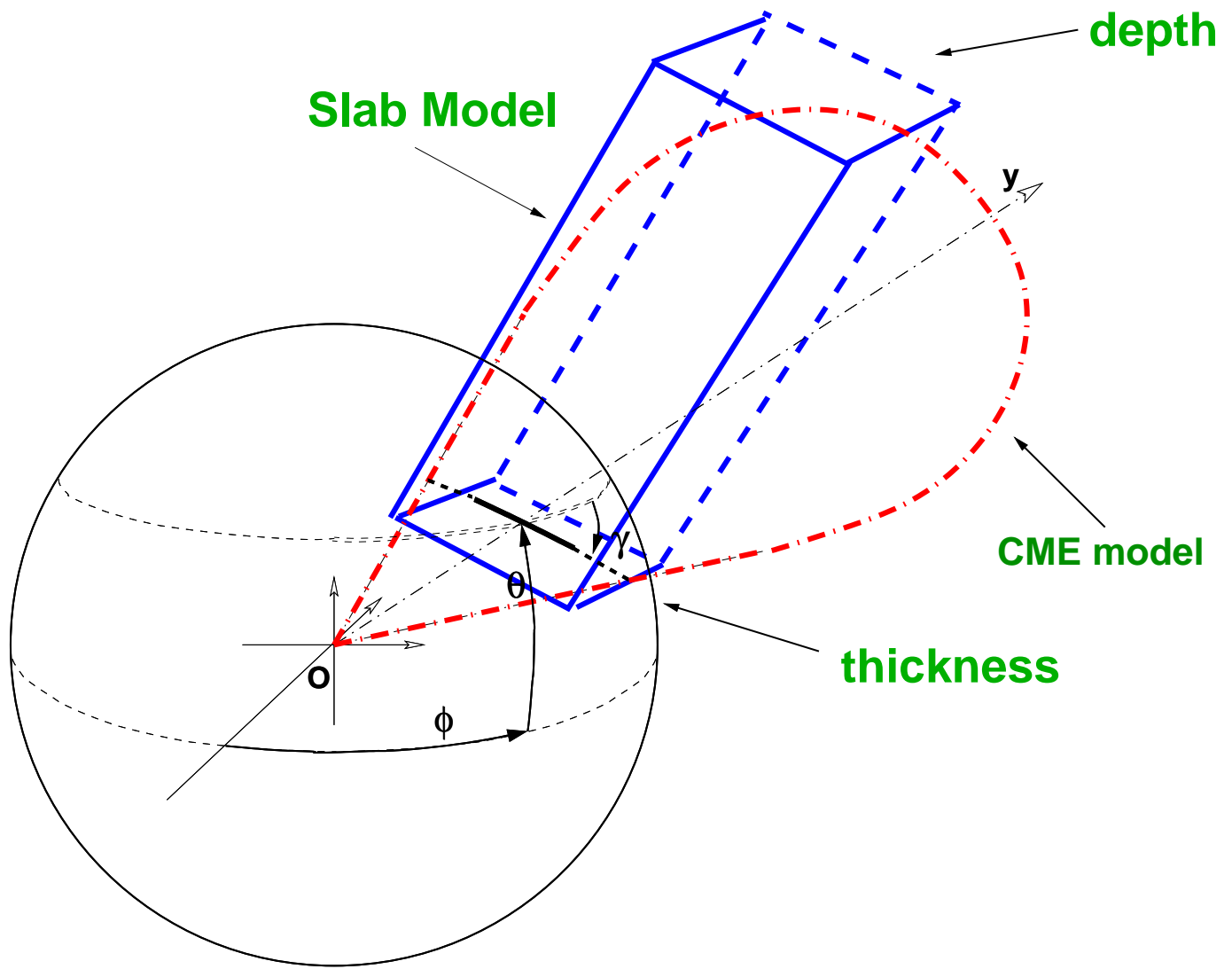}
\caption{Schematic of the 3D  models
used in the modeling of the WL CME  (red) and ray (blue).}
\label{cartoon}
\end{figure*}

Here, we model the observed CME  with a spherical shell attached to two  conical legs, a
so-called ``croissant'' model.  The spherical shell is meant to
reproduce the upper part of the CME, whereas the conical legs are an
approximation for the CME's lower sections, underneath the
spherical shell.   Such a model emulates
the outline of a flux-rope, widely accepted
as the magnetic structure of CMEs.
The model is formulated in terms of  the following free parameters: position
(longitude ($\theta$), latitude ($\phi$) of the legs on the solar surface ; tilt ($\gamma$) of the model baseline with respect to the equator; height above the 
solar surface  and aspect ratio (=height/radius)
of the spherical shell. The model is radial, i.e. it lies vertical to
the solar surface. A schematic of the model along with its free parameters
can be found in Figure \ref{cartoon}.

\begin{figure*}[p]
\includegraphics[scale=.45]{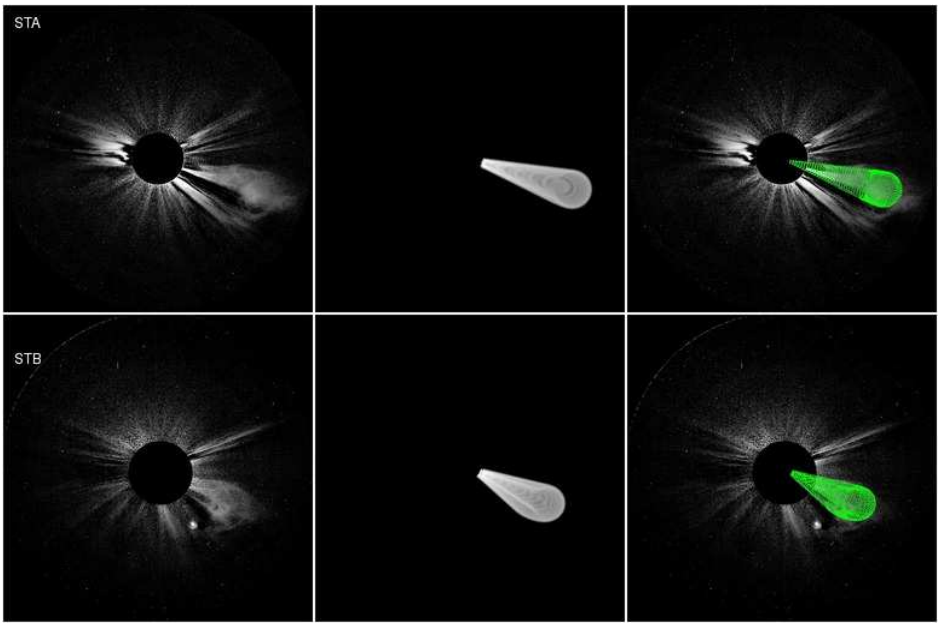}
\caption{3D geometrical modeling of the CME observations
of STA/COR2 (upper panel) and of STB/COR2 (lower panel). The left column shows the
TB images for each spacecraft, where a pre-event image 
was first subtracted. The middle column
shows synthetic TB images of the CME from our 3D modeling whereas
the right column shows overlays of the CME wireframe(red)
on the corresponding TB images. Observations taken
at 14:52:20 UT for STA and at 14:52:47 for STB.}
\label{cmemod}
\end{figure*}

The model parameters were varied until we achieved a visually
satisfactory agreement between the observed CME (in TB) and the
wireframe projection of the model was achieved {\it simultaneously}
for STA and STB (see left and right columns of Figure \ref{cmemod}
respectively).  We note that the model reproduces fairly well both the
body and the legs of the CME. We plot in the middle column of
Figure~\ref{cmemod} synthetic TB images from the CME model for STA and
STB to illustrate the model fitting further .   A uniform density
  of ${10}^{6} {cm}^{-3}$ was assumed; its precise value does not
  affect the purely geometrical analysis of this Section. Nevertheless
  the deduced geometrical information here will be used to
  determine the ray density in the following Section.  The figure shows
that the model appears to reproduce some of the internal
structure of the CME.  The WL renderings from the model show some fine
structure within the CME which is also present in the actual
observations.

\begin{figure*}[p]
\includegraphics[scale=.65]{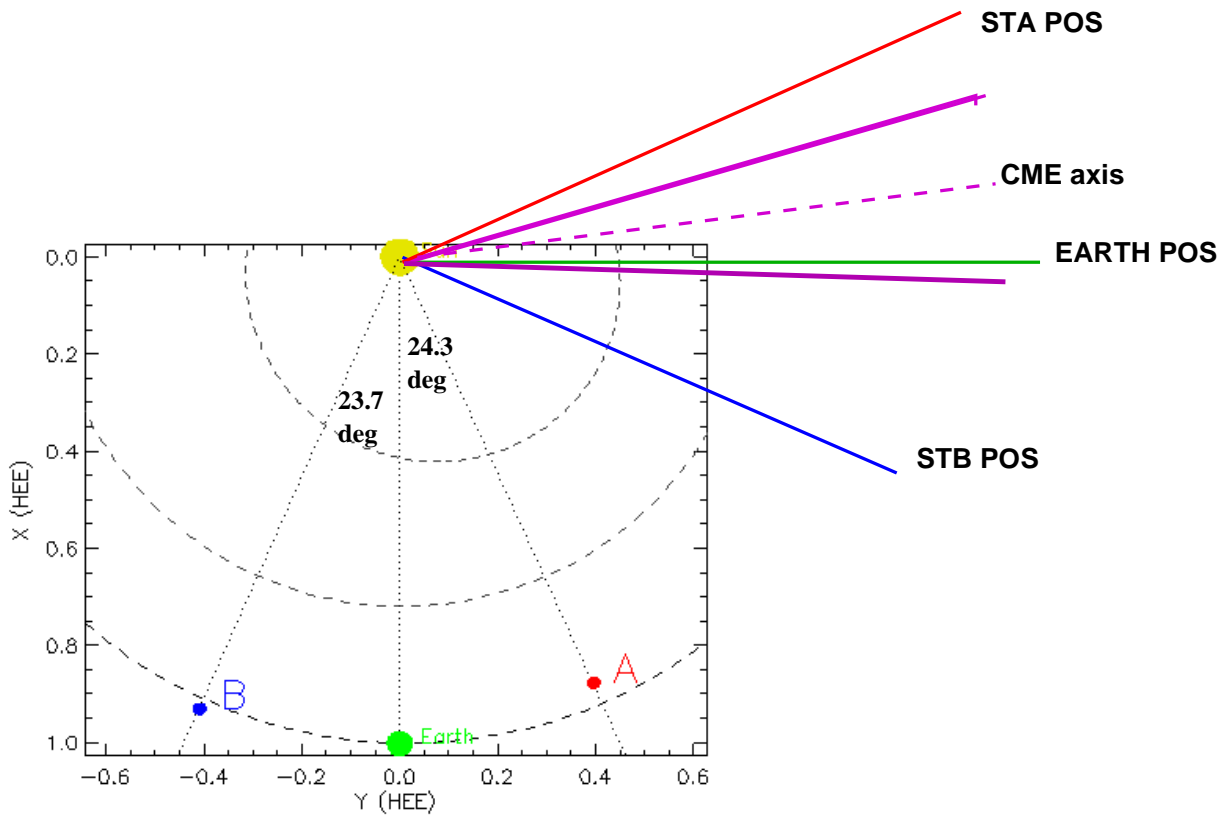}
\caption{Schematic of the orientation (dashed magenta line) and
of the angular width of the CME (solid magenta lines) as determined
from the 3D modeling of Section 3. The plot also supplies the 
relative positions of STB, Earth and ST as well as the corresponding
POS (blue, green and red solid lines respectively) during 9 April 2008.}
\label{where}
\end{figure*}

To estimate uncertainties in the model parameters we perturbed each
parameter of the "best-fit" model shown in Figure \ref{cmemod} by
keeping the remaining parameters frozen.  Each parameter was varied
until a visually unacceptable solution (i.e., projections) was found; this
supplied a measure of the uncertainty for each parameter.  The fitting
procedure was purely visual. Namely, we derived a set of parameters that give the best
visual correspondance between the model and the two-viewpoint observations.
The
"best-fit" parameters along with their uncertainties of the CME model
are given in Table 1. We note that the uncertainties are quite small,
which underlines the importance of using multi-view point data to
tightly constrain the CME model. 
This is particularly true for the CME
longitude which indicates the direction along which the CME
propagates.  Given the Carrington longitude of STA was $\approx
121^{\circ}$ on this day, the deduced Carrington
longitude of the CME ($\approx 194^{\circ}$) implies that the CME was
about $17 ^{\circ}$ away from the STA POS and therefore almost $11
^{\circ}$ and $67 ^{\circ}$ degrees from the POS of SOHO and STB
respectively.  Therefore the CME
was mainly an Earth and STA event. A schematic with 
the orientation and width of the fitted CME along with the locations
of STB, Earth and STA and their corresponding POS can be found in Figure
\ref{where}.

We then performed 3D geometrical modeling of the COR2A and C2 ray
observations.  We used again the Thernisien et al. (2006, 2009)
approach but with a different 3D geometrical model.  
A rectangular slab was
used instead. 
This is a reasonable approximation for the ray shape
since in both viewpoints it looks like an elongated with an almost constant
cross-sectional area surface. The standard theory of solar eruptions
(e.g., Forbes 2000) predicts  post-CME current sheet shapes
resembling rectangular slabs.
Our ray model was formulated in terms of the
longitude and latitude of the slab axis, and the slab width and
depth. A  schematic of the model along with its free parameters
can be found in Figure \ref{cartoon}.
Supposing for the moment that the ray represents a CS then its
depth would correspond to the longitudinal extent of the post-CME
arcade (along the erupting neutral line) whereas the ray width would
correspond to the thickness of the current layer where opposing
magnetic fields from both sides of the neutral line come into a
diffusion region.  For a ray viewed almost edge-on its depth will be
comparable to the LOS extent. Since the ray width and
depth are similar, the slab has a nearly square projection so the tilt becomes ambiguous (i.e.,
various tilts can fit the data equally well).  We therefore ``fixed'' the 
ray tilt to the "best-fit" tilt of the CME and derived a width and depth shown in Table~1.

\begin{figure*}[p]
\includegraphics[scale=.35]{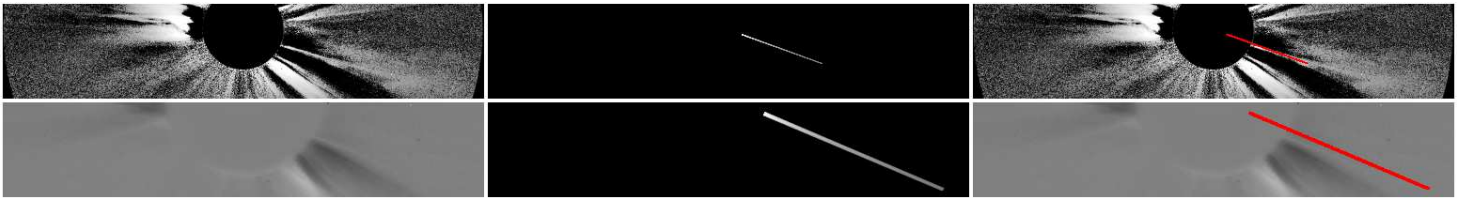}
\caption{3D geometrical modeling of the ray observations
of  STA/COR2 (upper panel) and of
LASCO/C2 (lower panel).The left column shows
TB images for each spacecraft, where a pre-event image 
was first subtracted. The middle column
shows TB images of the ray from our 3D modeling whereas
the right column shows overlays of the rectangular slab model wireframe(red)
on the corresponding TB images. Times of the observation:
17:52:20 UT for STA/COR2 and 17:50:04 for LASCO/C2.}
\label{csmod}
\end{figure*}

\begin{figure*}[p]
\includegraphics[scale=.35]{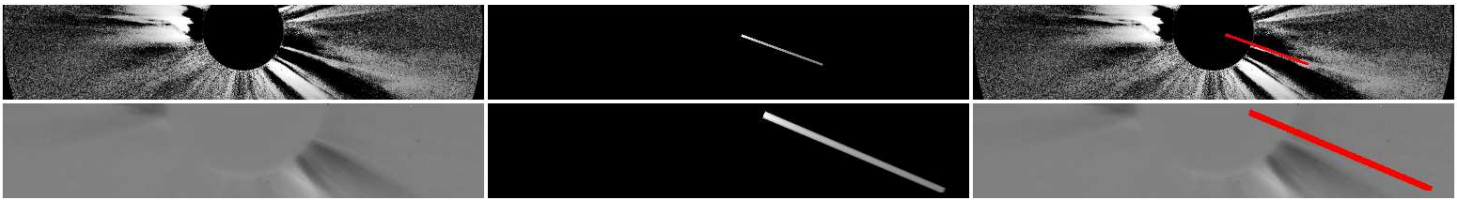}
\caption{Same as in Fig. \ref{csmod} but for a ray model with twice
the depth and the thickness of that of Fig. \ref{csmod}.}
\label{csmod1}
\end{figure*}

Figure~\ref{csmod} contains our ``best-fit'' model
of the ray.  As done for the CME
the "best-fit" model was determined
by visual comparison between the model predictions
and the actual multi-viewpoint observations.
As can be seen by this Figure our model
reproduces the ray
orientation and projected size in both viewpoints.
The uncertainties were derived  the same way as done
for the CME.  The ray parameters along their  uncertainties
are given in Table 1. 
Again, the simultaneous fitting of the two viewpoints
supplied strong constraints on the model parameters 
with the exception of the ray tilt angle due to its symmetry.
For instance we note that  increasing the ray depth and width
by a factor of two we get unacceptable solutions (Figure
\ref{csmod1}); the ray is too wide in the SOHO view.
Two important remarks can be now made.
First, the ray has a rather small width 
(0.05 $R_{\odot}$)
and depth (0.15 $R_{\odot}$) 
which
represent a small fraction of the solar radius. This
can be contrasted with the large radius of the CME (e.g., around
1.8 $R_{\odot}$ at 14:52 UT).
Second, the location of the ray (i.e, longitude and latitude)
is not very far off from the location of the CME.
For instance the CS central longitude differs by only   $3^\circ$
from that of the CME. This small difference is almost
within the error bars and certainly much smaller
than the CME width of $\approx 18^{\circ}$
This  means that the ray is almost perfectly aligned
with the CME and lies close to its center. Indeed the ratio between
the CME to ray volumes is $\approx$ 50, meaning
that the ray represented only a small fraction
of the CME volume.
We will discuss the important implications of the above findings
in Section \ref{disc}.

\section{Determination of the  Ray Density Structure}\label{den}

\begin{table}
\caption{CME and ray fitting results.}             
\label{table:1}      
\centering                          
\begin{tabular}{c c c}        
\hline\hline                 
   & CME & CS \\    
\hline                      
Carrington longitude($^\circ$) & 194 $\pm$ 5   & 197 $\pm$ 5  \\     
Carrington latitude($^\circ$)  & -17 $\pm$ 3 & -20 $\pm$ 3    \\
tilt($^\circ$) & 4 $\pm$ 7 & 4  $\pm$ 7     \\
aspect ratio & 0.16 $\pm$ 0.03 & N/A    \\
angular width($^\circ$) & 18 $\pm$ 4 & N/A    \\
depth ($R_{\odot}$)  &  N/A  &0.15  $\pm 0.08$   \\
thickness ($R_{\odot}$)  &  N/A  &0.05 $\pm 0.03$  \\
density $(\mathrm{{cm}^{-3}})$ &  N/A           & 3$\times 10^{7}$  $ \pm 5.3 \times {10}^{6}$ \\
\hline                                   
\end{tabular}
\end{table}

Since the WL intensities in coronagraph images are proportional
to the LOS integral of the electron
density (e.g., Billings 1966) we used our observations
to infer the density structure of the ray.
The common assumption made
in such determinations (e.g., Poland et al. 1981;
Vourlidas et al. 2000; Vr{\v s}nak  et al., 2009) is that the structure of interest
lies in the POS of the corresponding instrument.
Then, local densities are inferred by assuming
a LOS extent; a common assumption made is that 
it  is proportional
to the projected width of the structure.

Our measurements of the 3D shape and location
of the ray allowed to relax the above two
assumptions by supplying the distance
of the ray from the STA POS ($\approx 15^\circ$).
Since the ray orientation is degenerate
the LOS extent of the ray can be in principle anywhere between
the calculated ray thickness (0.05 $R_{\odot}$) and depth (0.15 $R_{\odot}$).
We make the assumption it is 0.10 $R_{\odot}$, i.e.
the median value of the above interval. Therefore, the LOS extent uncertainty
is $\pm 0.05 R_{\odot}$,  which translates into 50$\%$  density uncertainty. 
In such a way we were able to determine "true"
densities.

The determination of the ray densities involved the following steps.
First, we created an excess mass image
of the STA ray for the TB observation 18:07:54 UT UT by subtracting a pre-event image 
(taken at 09:22:54 UT). 
Both images were fully calibrated. 
The conversion from Mean Solar Brightness to electrons/cm$^{-2}$ 
was done following the standard assumptions of Thompson scattering (e.g., Vourlidas et al. (2000) for details) and the distance of the ray from the STA POS.

\begin{figure*}[p]
\includegraphics[scale=.8]{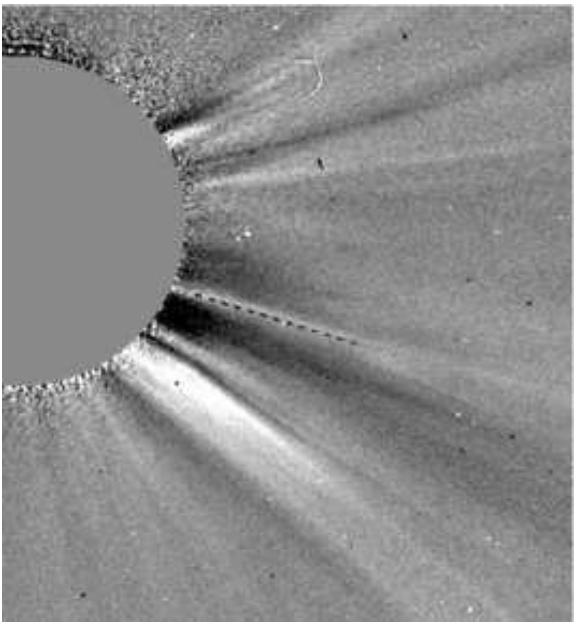}
\caption{Determining excess masses(densities) along the ray
(dashed line) 
from the COR2A observation at 17:52:20 UT. 
A pre-event frame taken at 09:22:00 UT
was subtracted. }
\label{trace}
\end{figure*}

The excess mass image is shown 
in Figure \ref{trace}. White (black) regions represent
places with excess (deficit) mass with respect
to the pre-event reference image.
Several points along the ray 
were then manually selected (dashed line in
Figure \ref{trace}). To enhance the ray signal
the mass measurements along the ray were binned
every 0.02 $R_{\odot}$. Then the excess mass  of the ray
was converted into a density by using its LOS extent
from our 3D modeling. We note here that 
the derived densities are likely lower limits to the true densities. 
If the CME removes significant part of the overlying streamer, 
then the pre-event subtraction will result in lower excess brightness than expected. 
This is insignificant in our case, since there is only a diffuse preexisting 
structure at the location of our ray and the observed depletions 
are minimal at the ray location (especially on the northward side).
Uncertainties due to photon counting statistics
can be safely ignored given the very large count rates we have
in WL at the heights of our observations (typically above thousand counts per pixel).
The uncertainty in the excess mass measurements
of the COR2 instruments is estimated to be 
20 $\%$.
The total uncertainty arises from the quadratic combination
of these uncertainties (mass determination and LOS depth) under the assumption they are independent (i.e., we took
the average of the quadratic sum of the individual uncertainties).

\begin{figure*}[p]
\includegraphics[scale=0.9]{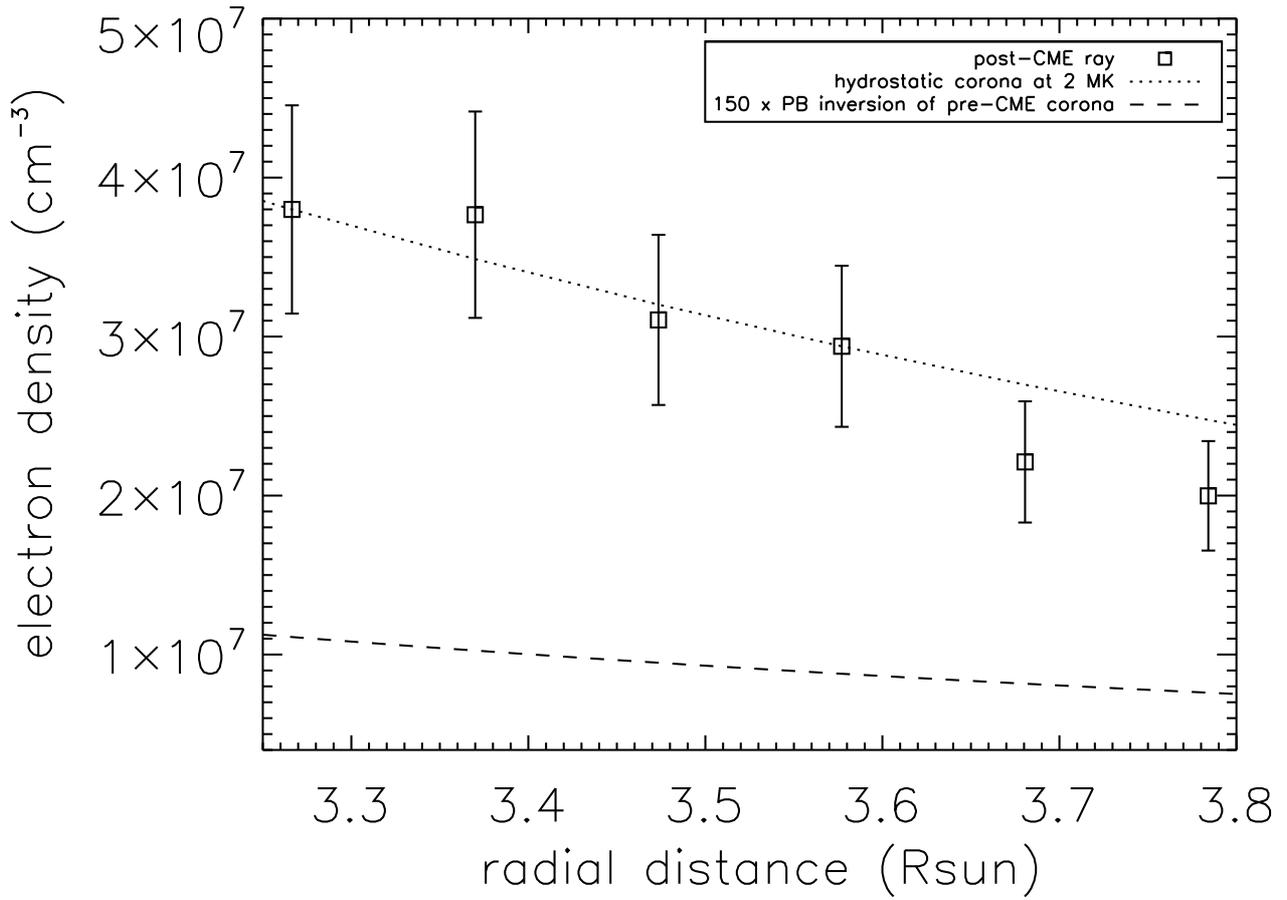}
\caption{Density along the ray as
a function  of  heliocentric distance (squares).
Density profile of the pre-CME corona on 9 April 2008, 10:06 UT from
LASCO C2 PB data inversion (dashes).
Density distributions for a hydrostatic isothermal corona
at 2 MK (dots).}
\label{ne}
\end{figure*}

The resulting density profile along the ray is given in Figure \ref{ne}.
Note also  that the   distances along the ray are also "true" in the sense
we used our 3D CS modeling to convert POS distances to heliocentric distances.
We plot densities only in the range 3.2-3.8 $R_{\odot}$  because the ray becomes too diffuse
at larger distances to be discerned from its surroundings.
Several remarks can now be made.
First, the derived density values are more than two orders of magnitude larger
than the densities  of the pre-CME corona
at the heights of our observations.  This can be seen in  Figure \ref{ne} where
we plot densities from the inversion of the LASCO C2 PB sequence taken 
at 10:06 on the day of the event. The sequence was taken
before the emergence of  our
CME into the C2 FOV and gives an estimate of the densities in the pre-CME corona. The pre-CME densities
were calculated at the same position angle as the ray densities in Figured~\ref{ne}.
The deduced ray
densities
are within the range  
$\approx {10}^{7}-{10}^{8} {cm}^{-3}$
determined
for several WL rays by  Vr{\v s}nak  et al. (2009).
Applying the same procedure to COR1 data, we found a ray density of $\approx 3\times{10}^{8} {cm}^{-3}$
at a distance of about 1.6 $R_{\odot}$. Again the ray emission becomes too faint at larger heights to be reliably measured.
Second, we note that the ray  density  does not substantially decrease with height
along its length (only a factor $\approx$ 2 over a distance of $\approx$ 
0.5 $R_{\odot}$). This implies reduced scales heights and therefore elevated temperatures within the ray.
Indeed, we can see from Figure \ref{ne} that the ray density drop-off
can be fairly well reproduced by a  hydrostatic density profile
at a  uniform temperature of 2 MK.
In doing this we  used a profile of the following form :
\begin{equation}
n(h)=n_{0}\exp(-h/H),
\end{equation}
where $h$ is the ray height above the solar surface, $n_{0}$
the density at the base of the ray and $H$ the isothermal scale-height (i.e., $\propto$ to temperature).
Empirical models of the quiet corona indicate temperatures of $\leq$ 1 MK at the heights of our observations (e.g., Withbroe 1988). 
We note here that the temperature
is inferred under the assumption of a hydrostatic equilibrium.
Such an assumption may not be fully justified for
post-CME rays which are often characterized by dynamic
phenomena such as plasma blobs.

\section{Discussion and Conclusions}\label{disc}
This paper presents the first multi-viewpoint observations
of a post-CME ray. Utilizing multiple view points
together with  detailed 3D geometrical modeling
supplied new important insight into
the 3D structure of a post-CME ray
and its relationship with the associated CME.
Our main findings are the following:

\begin{itemize}
\item The ray can be
fitted by a rectangular slab with small ($\approx 0.15 R_{\odot} $) depth
and  ($\approx 0.05 R_{\odot} $) thickness
\item The ray axis is almost aligned with that
of the CME
\item The ray lies within the CME volume and it occupies a much smaller volume than the CME 
\item The ray is characterized by elevated ($>$ 100)
densities with respect to the ambient corona
\item The ray density drop-off with height is consistent
with a high ($\approx$ 2~MK)  temperature, significantly hotter
than the ambient corona at the same heights. 
\end{itemize}

Our ray density and temperature findings are fully in
line with previous post-CME ray observations with UVCS and LASCO which
showed they are characterized by elevated densities and
temperatures. It is reasonable to expect such behavior within CSs.  The pinch-off of field lines resulting into a CS would
compress the plasma thereby increasing its density. Also significant
heating may occur in CSs since they are characterized by enhanced
current densities.

\begin{figure*}[p]
\includegraphics[scale=0.4]{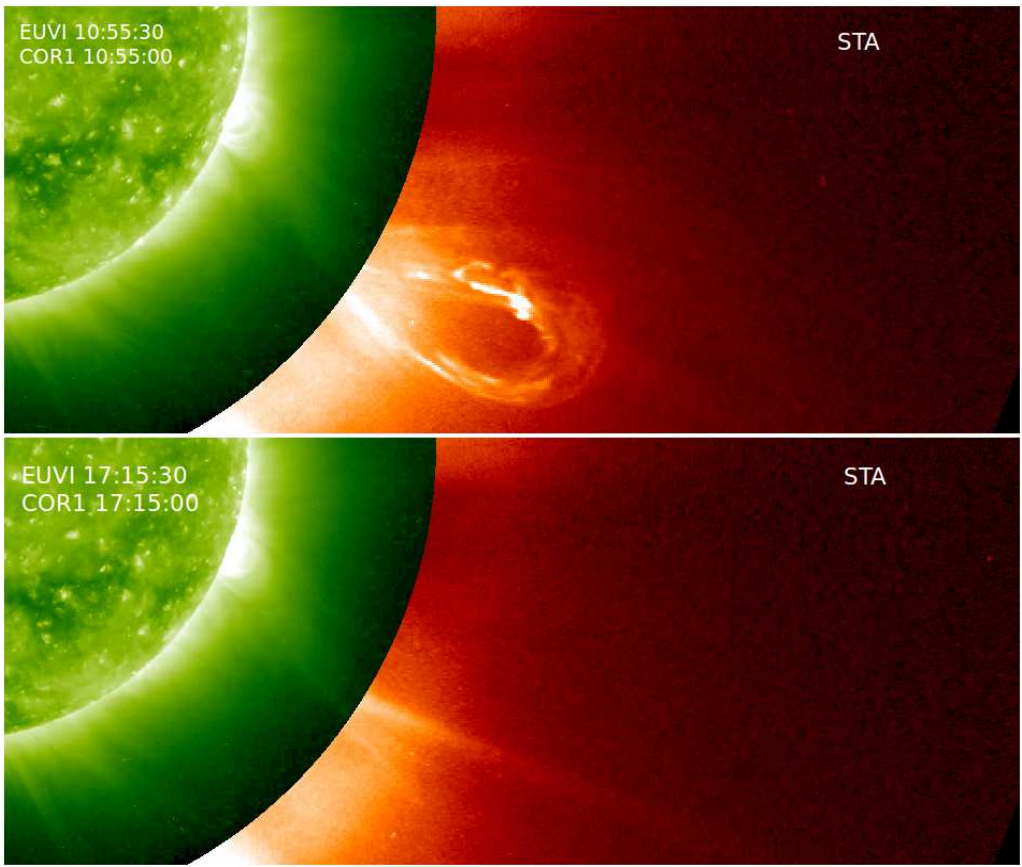}
\caption{Composite of EUVI-A 195 (inner FOV) and COR1-A (outer FOV) showing
the emerging CME (upper panel) and the resulting WL ray (lower panel). The temporal evolution is shown by movie1.mpg.}
\label{fest2}
\end{figure*}

\begin{figure*}[p]
\includegraphics[scale=.9]{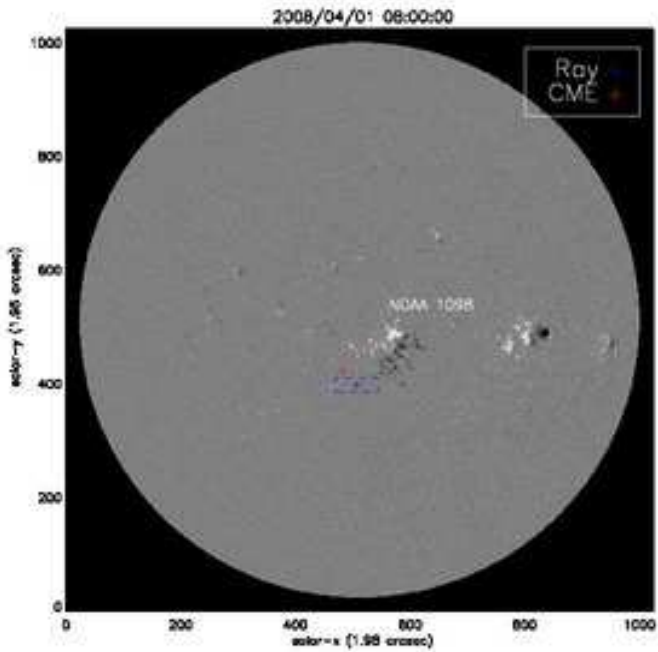}
\caption{Overlay of the CME (red cross) and
ray (blue cross) central longitudes and latitudes determined
from the 3D modeling of Section 3
on an MDI magnetogram taken  during 1 April 2008, when the host
AR (i.e., NOAA 1098)  was close to the central meridian as viewed from SOHO. 
The dashed blue  box
has a length/height equal to the depth/thickness of the WL ray as determined from our 3D modeling.
Its orientation cannot not be determined from our modeling (see the discussion of Section 3).}
\label{cs_mdi}
\end{figure*}

Undoubtedly the most important implications of our study arise from
the 3D analysis of the event.  We have determined the 3D shape of a
post-CME WL ray and deduced its width and depth. 
Figure~\ref{cs_mdi} contains an overlay of the CME and
the WL ray central longitudes and latitudes as determined from
the 3D modeling of Section 3 on  a MDI magnetogram taken on
1 April 2008, i.e. close to  the central meridian passage
of the host AR as viewed from SOHO. We first note
that both the CME (red cross) and WL ray (blue cross)
are significantly misplaced towards
the southwest direction with respect to the host AR.
This suggests  a non-radial evolution of the eruption in the inner corona and  verifies the visual
impressions of Section 2 (e.g., Figure \ref{fest2}).
Moreover, the 3D ray depth (equal to the vertical blue dashed
lines of Figure~\ref{cs_mdi}) is comparable to the length
of the neutral line of the host AR.
If the ray corresponds to a CS, we would expect its depth to be
comparable to the length of the erupting neutral line as seems to
be the case.  
However, the
deduced ray width ($\approx 0.05 R_{\odot}$) is still many orders of
magnitude bigger than the expectations of standard plasma theory.

We also found that the CME and the ray were almost
perfectly aligned. 
This supplies
strong evidence for a tight  and {\it causal}
relationship
between CME expansion and ray(CS) formation.
The ray(CS) should trail the CME  if it results from
the  pinch off
of magnetic field behind the eruption. The fact that the WL ray could represent
a post-CME CS could help removing the degeneracy in the ray
orientation discussed in Section 3 since they ray should lie in a plane
perpedicular to the plane of CME expansion. 
The fact that the ray and CME are not perfectly
aligned can be rather easily anticipated in terms
of the various re-arrangements (rotations, deflections etc)
taking place during the event.
Furthermore, according to the standard model for solar eruptions one expects that the
post-eruption CS in the inner
corona should lie above the post-flare loops which in turn are
straddling the neutral line of the erupting AR.
The fact that 
the candidate CS (i.e., the WL ray) is 
misplaced with respect to the erupting neutral line suggests the possibility
of a "broken" current sheet. This introduces a new, and possibly important, consideration for CME models.

\begin{figure*}[p]
\includegraphics[scale=.5]{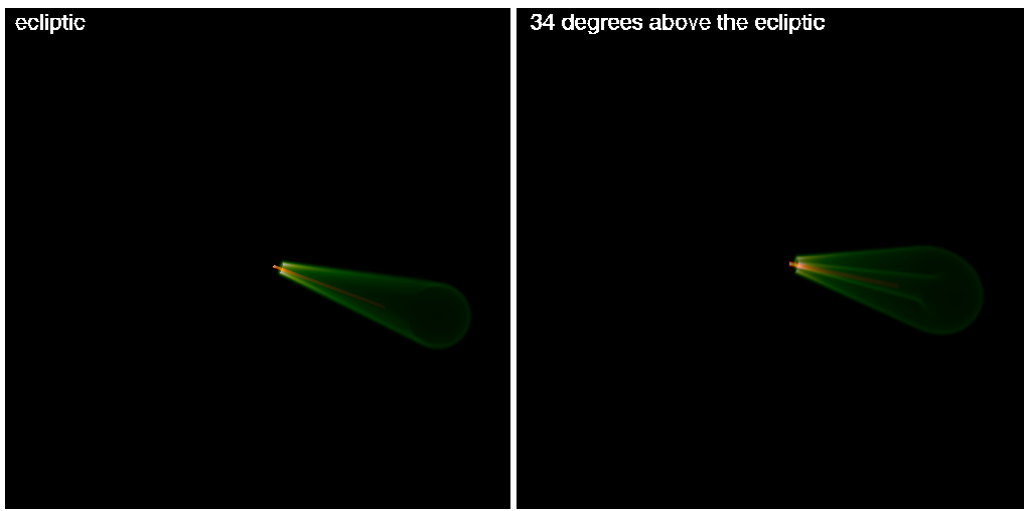}
\caption{Combined TB view of the CME (green)
and of the ray (orange). View from the ecliptic plane (left panel).
View  34 degrees above the ecliptic plane (right panel).
Both views are from  a distance
of 100 $R_{\odot}$.}
\label{render}
\end{figure*}

As a further visualization of the intimate
relationship between the ray and the CME
Figure \ref{render} presents WL renderings
of the ray and the CME from our modeling taken from
two different views: one from the ecliptic plane
and another   $34^{\circ}$ above the ecliptic.
The latter is a configuration which will be possible
with the planned Solar Orbiter mission.
What we note in this Figure is that the ray is much
smaller than the CME and it is located between its inner
legs. It seems therefore likely according to the standard
model for solar eruptions that once the CME leaves the FOV
its legs will pinch off and give rise to a ray (CS).

In summarizing, our study clearly
shows that the ray lies where it {\it should} be (aligned with the CME) 
and it is as big as it {\it should} be (depth similar to the erupting active region neutral line)
if it had been a CS.
Therefore, all our 3D findings give strong support
to the possibility that the observed post-CME ray is related to
a post-CME CS.

\acknowledgements

The SECCHI data used here were produced by an international consortium
of the Naval Research Laboratory (USA), Lockheed Martin Solar and
Astrophysics Lab (USA), NASA Goddard Space Flight Center (USA),
Rutherford Appleton Laboratory (UK), University of Birmingham (UK),
Max$-$Planck$-$Institut for Solar System Research (Germany), Centre
Spatiale de Li\`ege (Belgium), Institut d’ Optique Th\'eorique et
Applique\'e (France), and Institut d’Astrophysique Spatiale (France).
We thank the referee for useful comments which substancially impooved
the paper.


\newpage

\newpage

\newpage

\newpage

\newpage

\newpage

\newpage

\newpage

\newpage

\newpage

\newpage

\newpage

\end{document}